# Observation of two-dimensional coherent surface vector lattice solitons

M. Heinrich,[1] Y. V. Kartashov,[2] A. Szameit,[3] F. Dreisow,[1] R. Keil,[1] S. Nolte,[1] A. Tünnermann,[1] V. A. Vysloukh,[4] and L. Torner[2]

[1] *Institute of Applied Physics, Friedrich-Schiller-Universität Jena, Max-Wien-Platz 1, 07743 Jena, Germany*

[2] *ICFO-Institut de Ciencies Fotoniques, and Universitat Politecnica de Catalunya, Mediterranean Technology Park, 08860 Castelldefels (Barcelona), Spain*

[3] *Physics Department and Solid State Institute, Technion, 32000 Haifa, Israel*

[4] *Departamento de Fisica y Matematicas, Universidad de las Americas – Puebla, 72820, Puebla, Mexico*

We report the first experimental observation of vector surface solitons, which form at the edge and in the corner of two-dimensional laser-written waveguide arrays. These elliptically polarized vector states are composed of two orthogonally polarized components. They exist only above a power threshold and bifurcate from scalar surface solitons. The components of a vector soliton may have substantially different degrees of localization in certain parameter ranges.

*OCIS codes: 190.0190, 190.6135*

Solitons emerging from nonlinear interactions are a ubiquitous physical phenomenon arising in many areas of physics. Usually solitons appear as scalar (i.e., single-field) entities. However, solitons may also exist as vector objects, where two components are locked together in order to generate a localized state, with so-called Manakov solitons being one classical example [1]. Such vector optical solitons introduce nontrivial new physical effects [2-5]. Vector solitons have been predicted not only in uniform materials, but also in discrete waveguide arrays [6-8]. Such vector solitons were observed in both one- (1D) [9] and two-dimensional (2D) geometries [10]. Discrete vector solitons can also appear as multi-gap entities [11,12].

A unique laboratory to explore new soliton phenomena is provided by finite truncated periodic geometries. The properties of solitons at interfaces between uniform and periodic media have much in common with those shown by surface solitons at interfaces between natural materials, as predicted in [13]. Scalar lattice surface solitons have been observed at the edges of one- [14] and two-dimensional [15-17] periodic lattices. However, lattice interfaces also support vector solitons. The properties of the simplest one-dimensional discrete vector solitons were analyzed in [18], while complex vector surface states involving components from different gaps were obtained in [19,20]. Nevertheless, to date surface vector solitons were not observed experimentally.

In this Letter we report the first experimental demonstration of 2D coherent surface vector solitons at the edge and in the corner of a femtosecond-laser written waveguide array. We explore and observe the formation of coherent 2D vector states, sustained by cross-phase modulation (XPM) coupling between orthogonally polarized components in the presence of four-wave mixing.

To describe theoretically the formation of surface vector solitons, we assume continuous-wave (CW) illumination and use the nonlinear Schrödinger equation for the dimensionless field amplitudes $q_\text{x}$ and $q_\text{y}$ of the two coherently interacting, orthogonally polarized waves propagating along the $z$ axis:

$$i\frac{\partial q_\text{x}}{\partial \xi} = -\frac{1}{2}\left(\frac{\partial^2 q_\text{x}}{\partial \eta^2} + \frac{\partial^2 q_\text{x}}{\partial \zeta^2}\right) - q_\text{x}\left(|q_\text{x}|^2 + \frac{2}{3}|q_\text{y}|^2\right) - \frac{1}{3}q_\text{x}^*q_\text{y}^2 - p_\text{x}R_\text{x}(\eta,\zeta)q_\text{x},$$
$$i\frac{\partial q_\text{y}}{\partial \xi} = -\frac{1}{2}\left(\frac{\partial^2 q_\text{y}}{\partial \eta^2} + \frac{\partial^2 q_\text{y}}{\partial \zeta^2}\right) - q_\text{y}\left(|q_\text{y}|^2 + \frac{2}{3}|q_\text{x}|^2\right) - \frac{1}{3}q_\text{y}^*q_\text{x}^2 - p_\text{y}R_\text{y}(\eta,\zeta)q_\text{y}.$$

(1)

Here, $\eta,\zeta$ are the normalized transverse coordinates $x,y$ and $\xi$ is the normalized longitudinal coordinate $z$. According to our experimental results, $q_\text{x}$ and $q_\text{y}$ observe slightly different refractive index profiles $R_\text{x}$ and $R_\text{y}$ with corresponding modulation depths $p_\text{x} \neq p_\text{y}$. We approximate the shapes of the individual waveguides as $\exp[-(\eta/d_\text{x})^2 - (\zeta/d_\text{y})^2] + (1/2)\exp[-(\eta/d_\text{x})^2 - (\zeta/ad_\text{y})^2]$, with $a = 2$ for $q_\text{x}$ and $a = 4$ for $q_\text{y}$ respectively. The waveguide spacing in the square array is given by $d_\text{s}$. The second terms of the right sides of Eqs. (1) describe self-phase-modulation (SPM) and XPM, while the third terms introduce mismatch-free four-wave mixing (FWM) in isotropic fused silica.

Notice that a cumulative phase mismatch between $q_x$ and $q_y$ might still appear during propagation due to the difference of $p_x R_x$ and $p_y R_y$ profiles as well as because of nonlinear effects. We fixed $d_x = 0.8$, $d_y = 0.45$ and $d_s = 4$ in accordance with the transverse waveguide dimensions of $8 \times 4.5$ $\mu m^2$ and the spacing of $40$ $\mu m$. We set $p_x = 1.65$ and $p_y = 1.75$, while the value $p_x R_x = 2.5$ corresponds to an actual refractive index modulation depth of $2.8 \times 10^{-4}$. The quantity $U = U_x + U_y = \int\int_{-\infty}^{\infty}(|q_x|^2 + |q_y|^2)d\eta d\zeta$ corresponding to the total power, is conserved in Eq. (1).

Equations (1) admit scalar soliton solutions when one of the components is zero. Vector soliton solutions of Eq. (1) can have either the form $q_x = w_x \exp(ib\xi)$, $q_y = w_y \exp(ib\xi)$ (in-phase or out-of-phase components, i.e. linear polarization), or $q_x = w_x \exp(ib\xi)$, $q_y = iw_y \exp(ib\xi)$ (phase shift $\pi/2$ between components, i.e. elliptical polarization), where $w_x, w_y$ are real functions. As Eqs. (1) allow linearly polarized solutions only for $p_x R_x = p_y R_y$, we consider only elliptically polarized solutions. Their properties are summarized in Figs. 1 and 2. Exemplary profiles of surface vector solitons residing at the edge and in the corner of the array are shown in Fig. 2. The power $U_e$ of such solitons is a monotonically increasing function of its propagation constant $b$ (Fig. 1). All vector solitons exist above the cutoff $b_e^{co}$ and above a certain power threshold. While far from the cutoff, vector solitons are effectively confined to a single waveguide (middle column in Fig. 2), with decreasing $b$ the amplitude of $w_y$ decreases and the component expands widely across the array (left and right columns in Fig. 2). However, at the same time, $w_x$ remains well localized. Accordingly, the vector solitons components feature substantially different degrees of localization. This interesting effect occurs because the refractive index modulation depth observed by $w_y$ is slightly higher. In the scalar case, $w_y$ would exhibit a cutoff propagation constant $b_y^{co}$ which is higher than $b_x^{co}$ for $w_x$. Furthermore, its power $U_y$ would abruptly increase as $b$ exceeds $b_y^{co}$, which is typical for all surface solitons. In the vector case the already localized component $w_x$ induces an effective waveguide via XPM. The overall refractive index profile observed by $w_y$ is then given by $p_y R_y + (2/3)w_x^2$. Therefore, the power $U_y$ carried by $w_y$ vanishes when $b \to b_e^{co} \approx b_y^{co}$ (Fig. 1), and the shape of $w_y$ approaches that of a linear guided mode of the total index distribution $p_y R_y + (2/3)w_x^2$. At a certain power level $U_e$, the elliptically polarized vector soliton transforms into a scalar soliton of the $w_x$ component. In other words, for our set of parameters vector solitons bifurcate from the scalar soliton branch of $w_x$. With further decreasing $b$, the remaining $w_x$ component exhib-its all properties typical for scalar surface solitons, such as existence of a power threshold and broad expansion across the array near the corresponding cutoff propagation constant $b_\mathrm{x}^\mathrm{co} < b_\mathrm{e}^\mathrm{co}$ (Fig. 1). This behavior is shared by both edge and corner excitations. Note that the threshold for corner vector solitons is lower than that for edge solitons. Vector solitons are stable for sufficiently high powers and they are predicted to preserve their structure on propagation even when they are strongly perturbed.

In order to experimentally study surface vector solitons, we used the femtosecond laser direct-writing technique [21] to create a $7 \times 7$ square lattice of waveguides in a fused silica sample. In order to allow light to spread far across the array in the linear case within the given sample length of 105 mm as well as to form well localized solitons in the accessible power range, a waveguide spacing of 40 $\mu$m was chosen. For the excitation, we used a Ti: sapphire laser system (Spectra Tsunami/Spitfire), delivering 200 fs pulses at 800 nm with a repetition rate of 1 kHz. A beam splitter was used to divide the power equally; after rotating the polarization of one part using a half wave plate, another splitter was employed to reunite the beam. A manual delay line was utilized to compensate the different optical path lengths and ensure exact overlap of the two orthogonally polarized pulses. Light was injected into the respective lattice sites by a microscope objective ($2.5\times$, NA $= 0.07$), while another objective ($4\times$, NA $= 0.1$) was used to image the sample end facet onto a CCD camera. Specific polarization components were selected by a polarizer to ensure a separate monitoring.

Our observations of vector solitons at the edge and corner of the array are presented in the first and second row of Figs. 3 and 4, respectively. The measured linear diffraction patterns for both polarizations are shown in the first columns. One can see the repulsive force of the surface so that the beam penetrates deep into the array away from the surface. The second columns depict the uncoupled patterns for 500 kW pulse peak power, when the other respective component is switched off. A nonlinear modification of the intensity distributions is clearly visible, although the applied pulse power is not high enough to achieve localization. This picture changes when the two components interact by cross-phase-modulation and four-wave-mixing, as shown in the third columns. Although the same pulse peak power is applied as in the second column there is strong localization for both components, which indicates the energy exchange between both beams. This is a clear signature of the formation of a vector solitons state. In order to compare our experimental data with the numerics, in the bottom rows of Figs. 3 and 4 respective simulations of the $q_y$ field, corresponding to the thresholdless component, are shown. Note the excellent agreement between the experimental observations and the theoretical predictions.

In conclusion, we experimentally demonstrated the existence of stable, elliptically polarized surface vector solitons in two-dimensional femtosecond-laser-written waveguide arrays. They bifurcate from scalar surface solitons at a certain power threshold and can feature substantially different degrees of localization of the involved components.

The authors acknowledge support by the Deutsche Forschungsgemeinschaft (Research Unit 532 and Leibniz program), and the Government of Spain (grant TEC2005-07815 and Ramon-y-Cajal program).

# References with titles

# References without titles

# Figure captions

Figure 1.  Power versus propagation constant for (a) edge and (b) corner vector solitons. Circles correspond to the soliton profiles shown in Fig. 2.

Figure 2 (color online).  Profiles of surface vector solitons with $b = 0.616$ (left column), $b = 0.830$ (middle column), and $b = 0.598$ (right column). Left and middle columns correspond to edge solitons, while right column corresponds to a corner soliton. The top row shows $w_x$, while $w_y$ is shown in the bottom row. In all plots, white dashed lines indicate the position of the interfaces.

Figure 3 (color online).  Experimental output pattern for the $q_x$ component (first row) and the $q_y$ component (second row) for excitation of an edge waveguide. According simulations for $q_y$ are depicted in the third row. In column (a), the linear propagation patterns are shown. The nonlinear propagation at 500 kW peak power with no interaction between the components is shown in column (b), while in column (c) the components interact via XPM and FWM.

Figure 4 (color online).  Experimental output pattern for the $q_x$ component (first row) and the $q_y$ component (second row) for excitation of a corner waveguide. The arrangement of subfigures corresponds to Fig. 3.

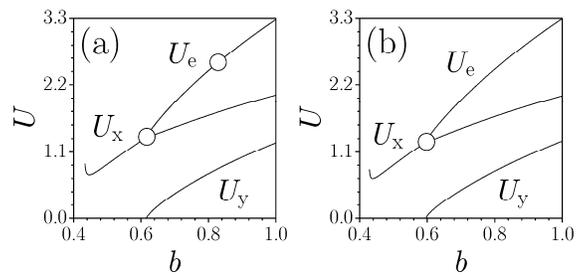

Figure 1.    Power versus propagation constant for (a) edge and (b) corner vector solitons. Circles correspond to the soliton profiles shown in Fig. 2.

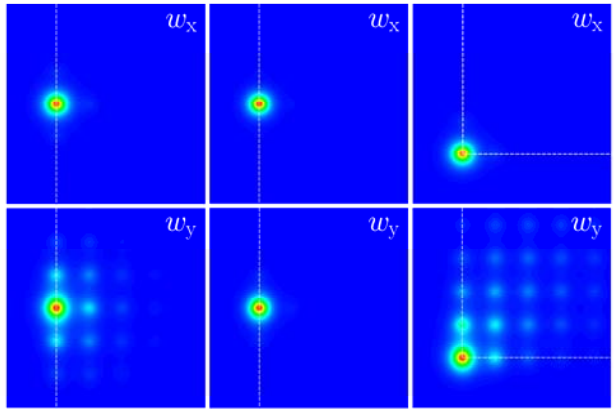

Figure 2 (color online).   Profiles of vector surface solitons with $b = 0.616$ (left column), $b = 0.830$ (middle column), and $b = 0.598$ (right column). Left and middle columns correspond to edge solitons, while right column corresponds to a corner soliton. The top row shows $w_\mathrm{x}$, while $w_\mathrm{y}$ is shown in the bottom row. In all plots, white dashed lines indicate the position of the interfaces.

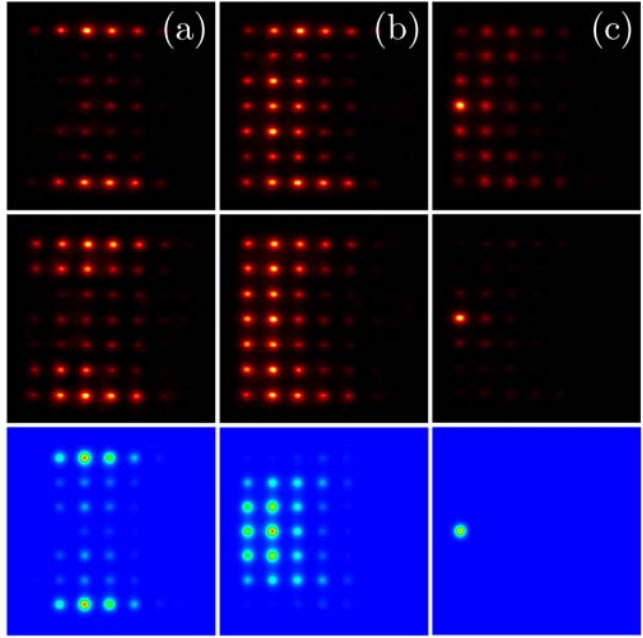

Figure 3 (color online). Experimental output pattern for the $q_x$ component (first row) and the $q_y$ component (second row) for excitation of an edge waveguide. According simulations for $q_y$ are depicted in the third row. In column (a), the linear propagation patterns are shown. The nonlinear propagation at 500 kW peak power with no interaction between the components is shown in column (b), while in column (c) the components interact via XPM and FWM.

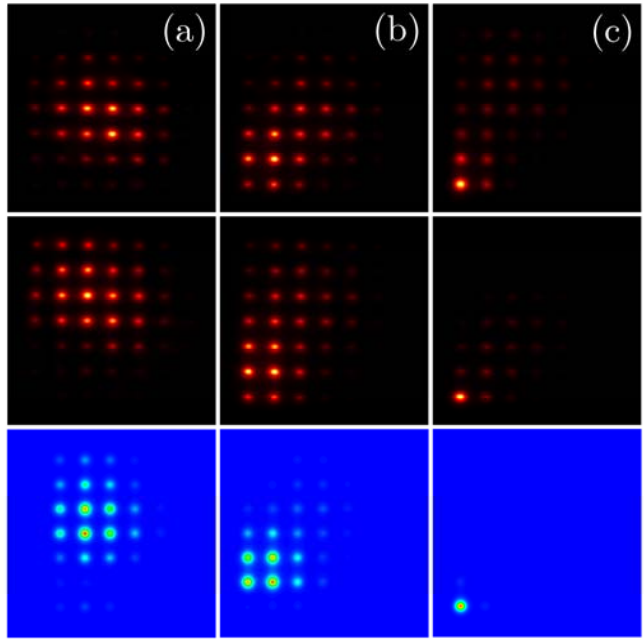

Figure 4 (color online). Experimental output pattern for the $q_x$ component (first row) and the $q_y$ component (second row) for excitation of a corner waveguide. The arrangement of subfigures corresponds to Fig. 3.